\documentclass[aps,prl,superscriptaddress,reprint]{revtex4-1}%
\usepackage{epsfig,pslatex,latexsym,times,amssymb,amsmath,graphicx}
\usepackage{bm, float, lipsum}
\usepackage{amsmath}
\usepackage{amsfonts}
\usepackage{amssymb}
\usepackage{mathrsfs}
\usepackage{color}
\usepackage{hyperref}
\usepackage{graphicx}%
\providecommand{\U}[1]{\protect\rule{.1in}{.1in}}
\begin{document}
\author{N. J. Harmon}
\email{nicholas-harmon@uiowa.edu} 
\author{M. E. Flatt\'e}
\affiliation{Department of Physics and Astronomy and Optical Science and Technology Center, University of Iowa, Iowa City, Iowa
52242, USA}
\date{\today}
\title{Distinguishing Spin Relaxation Mechanisms in Organic Semiconductors}
\begin{abstract}
A theory  is introduced for spin relaxation and spin diffusion of hopping carriers in a disordered system. For disorder described by a distribution of waiting times between hops ({\it e.g.} from multiple traps, site-energy disorder and/or positional disorder) the dominant spin relaxation mechanisms in organic semiconductors (hyperfine, hopping-induced spin-orbit, and intra-site spin relaxation)
each produce different characteristic spin relaxation and spin diffusion dependences on temperature. The resulting unique experimental signatures predicted by the theory for each mechanism  in organic semiconductors provide a prescription for determining the dominant spin relaxation mechanism.\end{abstract}
\maketitle	

Spintronics in organic semiconductors \cite{Vardeny2010} provides dramatically different regimes than are common for inorganic semiconductors\cite{Awschalom2002,Awschalom2007} or metals\cite{Ziese2001,Maekawa2006}, due to the very low mobilities and conductivities of organic materials, as well as the ubiquity of constituents with low atomic number (and thus often weak spin-orbit coupling (SOC) and long spin relaxation times). A major focus in organic spintronics has been on organic spin valves \cite{Xiong2004, Pramanik2007, Grunewald2009, Drew2009, Dediu2010}, within which spin polarized carriers are injected from one magnetic source contact into a non-magnetic organic spacer layer which they traverse before being collected by a magnetic drain contact. 
If the spins maintain their polarization over the length of the spacer region, then parallel and anti-parallel electrode magnetizations produce different resistances and lead to  spin-valve phenomenology.
The spin-valve effect requires the preservation of spin within the spacer layer;
therefore the reported long spin lifetimes in organic semiconductors increase organic spin valves appeal.
Any spin-based application beyond a spin-valve should also require an understanding of the rate of decay of the spin polarization in organic semiconductors.
The dominant spin relaxation mechanism is controversial \cite{Pramanik2006, Pramanik2007, Bobbert2009, Nguyen2010, Yu2011, Schulz2011} but appears to be driven by either SOC or the hyperfine interaction (HFI) during hopping, or associated with an intra-site relaxation (ISR) process that may include spin interactions with phonon modes or the multiple nuclear fields at a molecular site (independent of hopping).
Only a few theoretical investigations \cite{Bobbert2009, Yu2011} have examined  spin relaxation including disorder, even though  the interplay between spin and charge dynamics is well-known for inorganic semiconductors\cite{Flatte2000b}.

Here we provide a unified theoretical description of spin lifetimes in organic systems from SOC, HFI or ISR, based on a continuous-time random-walk theory \cite{Scher1975, Bottger1985}, that predicts dramatically different spin lifetime dependencies on temperature as well as previously unrecognized analytic dependencies on other physical properties of the organic semiconductor; these results are summarized in the upper half of Table I.
We introduce disorder in two scenarios: (1) by allowing hopping carriers to be captured by traps at a single energy or (2) at a wide array of energies. The second scenario uses a hopping-time distribution that lacks a first moment (a so-called heavy-tailed wait-time distribution (WTD)) and also describes charge transport when energetic disorder of a certain type is present that produces time-dependent mobility  (so-called `dispersive' transport). These heavy-tailed WTDs, which at long times scale as $t^{-1-\alpha}$ where $\alpha$ is a characteristic parameter of the WTD, provide the phenomenology associated with each spin relaxation mechanism (SOC, HFI, or ISR) during dispersive transport.
Dramatic qualitative differences emerge between HFI and SOC spin relaxation times ($\tau_s$) and spin diffusion lengths ($\ell_s$), which positions our theory to predict experimental observations to differentiate between the two; {\it e.g.} our theory provides supporting evidence that SOC dominates in Alq$_3$ whereas HFI dominates in MEH-PPV. 
In addition, HFI, SOC, and ISR are incorporated simultaneously \emph{within a single theory}, which illuminates  cross-over behavior from one mechanism to another. 
Our results for low disorder, reported in the non-dispersive section of Table I, agree with previously-obtained analytic results for SOC \cite{Yu2011} and HFI \cite{Yaouanc2011} spin dynamics. We further agree with Ref.~\onlinecite{Bobbert2009}'s result for the  spin diffusion length in the fast hopping HFI dispersive regime for the special case of $\alpha = 1/2$.
By varying temperature, HFI strength, and SOC, the entries of Table I change in a unique manner; our theory will aid experimental efforts to distinguish which relaxation route occurs.
As our theory uses an arbitrary WTD, it is applicable to a wide variety of other types of disorder and materials, such as spin transport in amorphous silicon or colloidal quantum dot films.

\begin{table}[h]
\begin{tabular}{| c | c | c | c | c |}
\multicolumn{5}{c}{}\\
\hline
dispersive	& ISR & HFI (fast) & HFI (slow) & SOC  \\ \hline
$1/\tau_s$	& $\Gamma$ & $\sim a$ 				      & $\sim k_0$  & $\sim k_0 \gamma^{2/ \alpha}$ \\
$\ell_s/\overline{r}$	& $\sim$\big($\frac{k_0}{\Gamma}\big)^{\alpha/2}$	& $\sim$\big($\frac{k_0}{a}\big)^{\alpha/2}$& $\sim 1$  &  $\sim \frac{1}{\gamma}$ \\
\hline
non-dispersive	& ISR & HFI (fast)  & HFI (slow) & SOC  \\ \hline
$1/\tau_s$& $\Gamma$	& $2 \frac{a^2}{k_0}$       & $\sim k_0$  & $k_0 \gamma^{2}/3$ \\
$\ell_s/\overline{r}$	& $ \sqrt{\frac{k_0}{\Gamma}}$	&$ \frac{k_0}{\sqrt{2}a}$& $\sim 1$  &  $\frac{\sqrt{3}}{\gamma}$ \\
\hline
\end{tabular}
\caption[]{Spin relaxation rate ($\tau_s^{-1}$) and spin diffusion length ($\ell_s$) dependencies on width of the Gaussian HFI distribution ($a$), SOC ($\gamma$), hopping rate ($k_0$), and ISR ($\Gamma$). 
$k_0$, $\Gamma$, and $\alpha$ depend on temperature, and thus allow access to different regimes in the same sample.  Deuteration affects $a$ and elemental composition affects $\gamma$. Fast hopping HFI  considers only  short-timescale spin relaxation/diffusion.
Spin decay in the dispersive regime is algebraic in time.}
\end{table}

\begin{figure}[ptbh]
 \begin{centering}
        \includegraphics[scale = 0.285,trim = 30 117 100 97, angle = -0,clip]{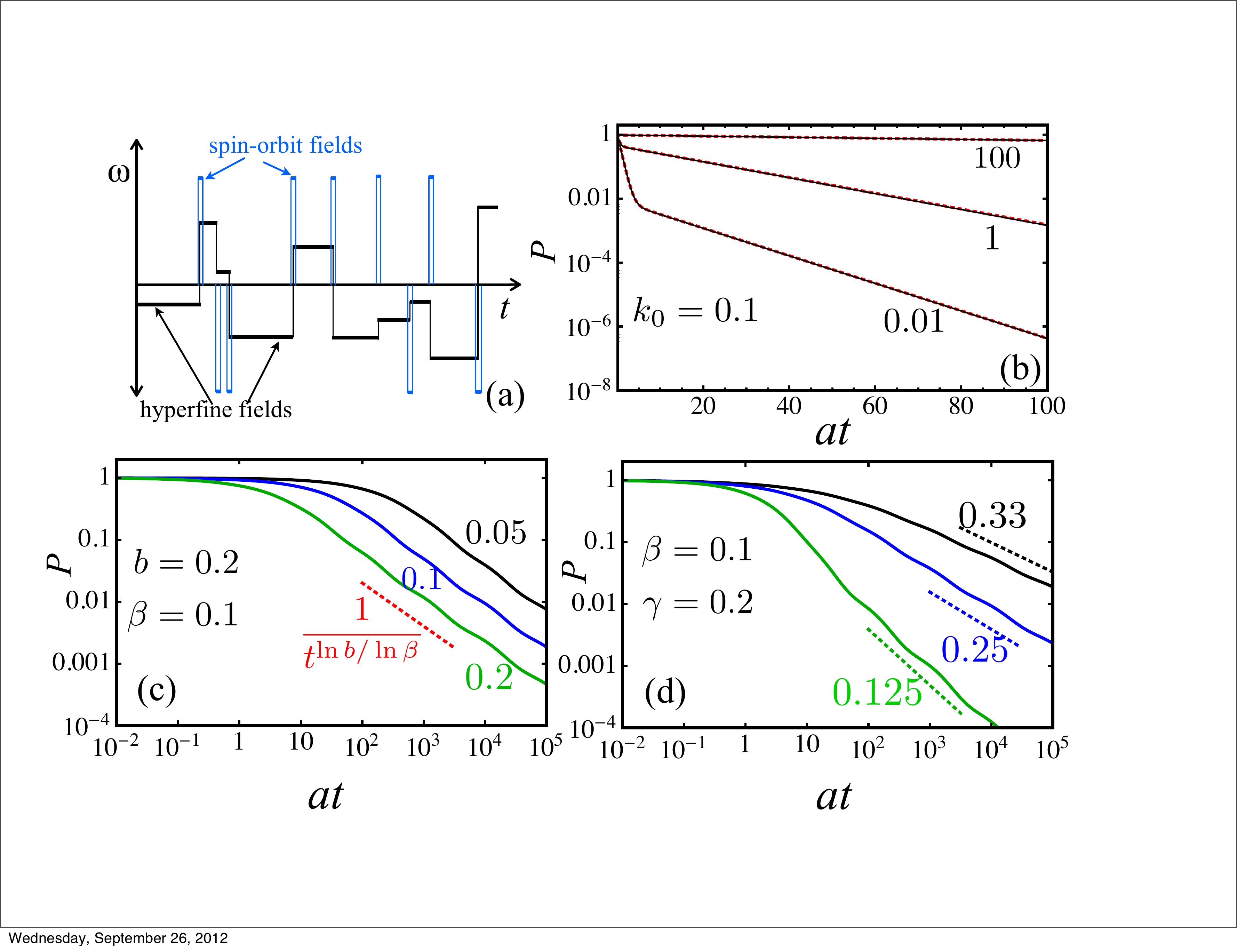}
        \caption[]
{(Color online) (a) Schematic of the time-dependent fields a single spin experiences from HFI and a constant ($\gamma$) SOC.
For (b), (c), and (d): $k$, the average hopping rate, is 100, $a=0$ and $\Gamma=0$ so these results are for SOC only.
(b) Polarization  as a function of time for explicit trap (black) and bi-exponential (red dotted) models. 
The rate of leaving a trap, $k_0$, for (a), $k_0 = 10$ and (b), $k_0 = 0.1$ where each pair is labeled by $k_1$, the rate to be captured by a trap.
Spin relaxation decreases as $k_1$ increases (i.e. as trap number increases). 
The two models are connected through the relations $b = k_1/(k+k_1)$ and $\beta = k_0/k$.
Multi-exponential model in (c) with varying SOC labels and in (d) with different $b$.  Slopes of long time power law dependence are indicated.
}\label{fig:SO1}
        \end{centering}
\end{figure}

\emph{Theory with general WTD} - 
The dynamics of a classical spin $\bm{S}$ in an arbitrary precessional field $\bm{\omega}$,   influenced by an ISR  process with a phenomenological rate $\Gamma$, is
\begin{equation}\label{eq:spinDiffEq}
\frac{d\bm{S}(t)}{dt} = \bm{\omega}  \times \bm{S}(t) - \Gamma \bm{S}(t) = (\bm{\Omega} -  \Gamma  \hat{\bm{1}})\cdot \bm{S}(t)= (\omega \hat{\bm{\Omega}} -  \Gamma  \hat{\bm{1}}) \cdot \bm{S}(t),
\end{equation}
where $\bm{\Omega}$ is a  skew-symmetric matrix \cite{supp}, and is solved by
$\bm{S}(t) = e^{-\Gamma t} e^{\bm{\Omega}t} \cdot \bm{S}_0 \equiv e^{-\Gamma t} \hat{\bold{R}}(t) \cdot \bm{S}_0,$
where $\bm{S}_0$ is the initial spin vector and 
\begin{equation}\label{eq:rotationMatrix}
\hat{\bold{R}}(t)  = \hat{\bm{1}}+ \sin\omega t \hat{\bm{\Omega}} + 2 \sin^2\frac{\omega t}{2} \hat{\bm{\Omega}}\cdot \hat{\bm{\Omega}}.
\end{equation}
Here we consider fields from two sources: a hyperfine field ($\bm{\omega}_{hf}$) and a spin-orbit field ($\bm{\gamma}$).
Spins at different sites which begin in the same polarization state experience different hyperfine fields, which causes partial dephasing of the spin ensemble \cite{Kubo1966}; polarization is further lost when the spins incoherently hop and experience different hyperfine fields at other sites \cite{Kehr1978}. Hopping also affects spin polarization through the spin-orbit interaction; spin-flips are possible as the carriers are not in pure spin states \cite{Tamborenea2007, Intronati2012, Yu2011}.
Figure \ref{fig:SO1}(a) schematically indicates the fields a spin might experience over a time interval due to hyperfine and spin-orbit fields. 
Lastly a spin at any site may interact with  phonon modes which leads to ISR \cite{Yu2008, Schulz2011}.

To describe the evolution of an ensemble, we simplify the single-spin dynamics by assuming:
1) hyperfine and spin-orbit rotations are independent and isotropic in space,
2) hyperfine rotations occur discontinuously at each hop \cite{Hayano1979} (as represented in Figure \ref{fig:SO1}(a)), 3) spin-orbit rotations of magnitude $\gamma$ (also treated phenomenologically) are independent of the duration of time spent at a site (they only occur at hops) and
4) ISR is independent of hopping.
Hyperfine fields are 
distributed according to a Gaussian distribution with width $a$ - a phenomenological parameter assumed to be on the order of 1 mT. The random direction of spin-orbit fields results from the random orientation of localizing molecular sites \cite{Yu2011}.

After averaging Eq. \ref{eq:rotationMatrix} over the hyperfine field distribution, 
\begin{eqnarray}
\hat{\bold{R}}_{hf}(t)  \equiv \langle \hat{\bold{R}}(t)  \rangle &=& 
 \left( {\begin{array}{ccc}
 R_x      & -R_{xy}   & 0 \\
 R_{xy} & R_x  & 0\\
  0         & 0      & R_z  \\
 \end{array} } \right),
\end{eqnarray}
whose elements are found in \cite{supp}. 
The spatially averaged spin-orbit rotation matrix is 
$\hat{\bold{R}}_{\gamma} = 
\hat{\bm{1}} - (4/3)
\sin^2(\gamma/2)
\hat{\bm{1}}$.
Typically, $\gamma$ is assumed to be a small angle such that $\hat{\bold{R}}_{\gamma} \approx 
\hat{\bm{1}} - \frac{1}{3}\gamma^2 \hat{\bm{1}}$. 

We now consider the spin ensemble to be hopping from site to site where the WTD is arbitrary; the time evolution of the spin polarization can be calculated and the spin relaxation time can be extracted  - numerically in general, but analytically in certain cases to be discussed.
Since the HFI operates between hops and the SOC affects spin at the hopping event, their rotation matrices enter into the polarization function in different ways which preclude interference effects. 
In this sense, HFI is analogous to the D'yakonov-Perel' \cite{Dyakonov1972} and SOC is analagous to the Elliott-Yafet \cite{Elliott1954, Yafet1963} spin relaxation mechanisms for inorganic semiconductors.

The rotation accrued during the wait-time is $\hat{\bold{R}}_s$, which is $\hat{\bold{R}}_{hf}$ when  HFI is present and $\hat{\bold{1}}$ when it is absent.
An angular rotation $\hat{\bold{R}}_h = \hat{\bold{R}}_{\gamma}$ occurs from the hopping event when SOC is present but is  $\hat{\bold{1}}$ when SOC is absent.
The third spin loss avenue, intra-site spin relaxation, simply appends $ \exp{(-\Gamma t)}$ to $\hat{\bold{R}}_s$ as seen in Eq. (\ref{eq:spinDiffEq}).
With these different factors in mind, we write the $\hat{\bold{R}}$ due to all single hops in the spin ensemble as
\begin{equation}
\hat{\bold{R}}_{1}(t)= \hat{\bold{R}}_h \int_0^{t} \hat{\bold{R}}_s(t-t')\Phi(t-t') e^{-\Gamma (t-t')}  \bigg( \hat{\bold{R}}_s(t') \psi(t') e^{-\Gamma t'}\bigg) dt'.
\end{equation}
The hopping rotation matrix is factored out of the integral since it is independent of time.
$\psi(t)$ is the WTD from which the wait-time $t$ is drawn. It appears once since one hop occurs. $\Phi(t)$ is the survival probability: the probability that the wait-time at a site exceeds $t$: $\Phi(t) = \int_t^{\infty} \psi(t')dt'$ \cite{Klafter2011}.
The quantities
$ \hat{\bold{R}}_{0}(t)\equiv \hat{\bold{R}}_{s}(t)  \Phi(t) e^{-\Gamma t}$ and $ \hat{\bold{R}}'_{0}(t)\equiv \hat{\bold{R}}_{s}(t) \psi(t) e^{-\Gamma t}$
permit the compact expression
$\hat{\bold{R}}_{1}(t) =  \hat{\bold{R}}_h \int_0^{t} \hat{\bold{R}}_0(t-t')  \hat{\bold{R}}'_0(t') dt'.$
The procedure can be continued indefinitely for an arbitrary number $l$ of hops yielding the  recursive equation
\begin{equation}\label{eq:recursion}
\hat{\bold{R}}_{l+1} (t)=  \hat{\bold{R}}_h \int_0^{t} \hat{\bold{R}}'_0(t-t') \hat{\bold{R}}_l(t') dt',
\end{equation}
which has the form of a convolution; 
$\tilde{\hat{\bold{R}}}_{l}(s ) =  \hat{\bold{R}}_h^{l} \tilde{\hat{\bold{R}}}_{0}(s + \Gamma) \tilde{\hat{\bold{R}}}'^{l}_{0}(s + \Gamma)$ in Laplace space.
We build the polarization function from the different number of random rotations \cite{Kehr1978, Belousov1979}:
\begin{equation}\label{eq:polarization}
\bold{P}(t) = \hat{\bold{P}}(t)\cdot \bold{S}_0 = \sum_{l=0}^{\infty}\hat{\bold{R}}_{l}(t)\cdot \bold{S}_0,
\end{equation}
where $\hat{\bold{R}}_{l}(t)\cdot \bold{S}_0$ is the polarization after $l$ jumps occurring between time 0 and $t$.
The polarization can  be calculated in Laplace space by summing the geometric series Eq. \ref{eq:polarization} to obtain our main equation:
\begin{equation}\label{eq:main}
\tilde{\bold{P}}(s) = \tilde{\hat{\bold{R}}}_{0}(s + \Gamma)
[\hat{\bm{1}} - \hat{\bold{R}}_h \tilde{\hat{\bold{R}}}'_{0}(s + \Gamma)]^{-1} \cdot \bold{S}_0.
\end{equation}
We have left the polarization as a vector since when an applied field is present, the rotation matrices need not be isotropic and the polarization in general decays with different longitudinal and transverse spin relaxation rates.

An exponential WTD, $\psi(t) = k \exp(-k t)$ characterized by an average hopping rate $k$, provides a clarifying example of the theory, as $\hat{\bold{R}}'_{0}(t) = k \hat{\bold{R}}_{0}(t)$ and the Laplace transform of $\hat{\bold{R}}_{0}(t)$ is especially simple:
$\int_0^{\infty}\hat{\bold{R}}_{0}(t) e^{-s t}dt = \tilde{\hat{\bold{R}}}_{s}(s+ \Gamma + k)$.
Eq. \ref{eq:main} becomes
\begin{equation}\label{eq:laplace}
\tilde{\bold{P}}(s) =\tilde{\hat{\bold{R}}}_s(s + \Gamma +  k)[\hat{\bm{1}} -k \hat{\bold{R}}_h \tilde{\hat{\bold{R}}}_s(s +\Gamma +  k)]^{-1} \cdot \bold{S}_0.
\end{equation}
With HFI present, the polarization cannot be solved analytically in the time domain except in special cases. This polarization function has been studied in the context of muon spin rotation\cite{Belousov1979, Smilga1994, Yaouanc2011}. 
For only SOC (and $\Gamma = 0$), the polarization in the time domain is calculated exactly to be
$P(t) = e^{-k \gamma^2 t/3}$, which is in qualitative agreement with Yu's recent determination \cite{Yu2011, Yu2012}.
If $\gamma = 0$, then $\tilde P(s) = 1/(s  + \Gamma)$ (i.e. $P(t) = \exp(-\Gamma t)$) can be shown from the relation between $\tilde\Phi$ and $\tilde\psi$.
Ref. \onlinecite{supp} contains a general expression for Eq. (\ref{eq:main}) in the absence of hyperfine fields. 

\emph{Traps: simple models} - 
Before examining systems with realistic WTDs, three instructional models are investigated that have analytic solutions.
The easiest model to study is that of a single trapping level. 
There are two ways to obtain nearly equivalent results: 1) incorporate explicit traps by forming a modified renewal equation in the spirit of Eq. (\ref{eq:main}) and 2) use a bi-exponential WTD to describe normal hopping and trap release.
The first approach was used in a limited manner by Kehr and Honig \cite{Kehr1978}. 
We have generalized their approach to include SOC but only report the results here in Figure \ref{fig:SO1} (b) (black lines) where analytic solutions for the polarization function are achievable if $a = 0$. 

The second approach accounts for traps by using the bi-exponential WTD:
$\psi(t) = (1+b)^{-1} k e^{-k t} +b\beta(1+b)^{-1} k e^{-\beta k t}$
with $0 < \{b, \beta\} < 1$.
We consider $b$ to be controlling the importance of traps. When $b$ is small, traps become inconsequential as seen from the fact that the single exponential WTD emerges.
The dimensionless quantity $\beta$ determines the reduction of the trap release rate compared to the normal hopping rate, $k$.
It is important to mention that spin relaxation results within the two-state trap model and bi-exponential WTD are always fit well with exponential decay times (aside from short time dynamics).
Figure \ref{fig:SO1}(b) shows an example of SOC-only spin relaxation (red-dotted lines) and compares the explicit and bi-exponential trapping models.
Since spin decay is produced by hops, long release times ($1/k_0$) promote long lived polarizations as shown in Figure \ref{fig:SO1}(b).
Likewise more traps (larger $k_1$) reduces overall number of hops and spins live longer.

Heavy-tailed WTDs have successfully been used to describe charge transport in disordered systems \cite{Scher1975}. 
The success of the bi-exponential model suggests an easy way to extend the model to a diverse array of trap frequencies.
We construct a multi-exponential WTD from the bi-exponential model as follows: we allow for more trap release times by taking $\beta \rightarrow \beta^j$ with $j$ being a positive integer.
Trapping probabilities are also generalized to $b \rightarrow b^j$.
The resulting WTD has been previously studied \cite{Shlesinger1981, Shlesinger1988, Scher1991}:
$\psi(t) = b^{-1}{(1-b)}\sum_{j=0}^{\infty} b^{j+1}\beta^j k e^{-\beta^j k t}.$
For the condition $b < \beta$, this WTD does not have an average wait-time which distinguishes it from the bi-exponential WTD. 
At long times, the algebraic form is found to be $\psi(t) \sim t^{-\alpha-1}$ where $\alpha = \ln b/\ln \beta$ \cite{Shlesinger1981,Klafter2011}.
Figure \ref{fig:SO1}(c, d) show different scenarios with this heavy-tailed WTD. Axes are log-log to emphasize the algebraic decay as opposed to the exponential decay in Figure \ref{fig:SO1}(b).

\emph{Mutiple-trapping} - We now examine a realistic WTD ---  one due to trapping levels distributed exponentially and for HFI and ISR as well as SOC.
It is now thought that hopping within an energetic distribution of hopping sites is much like what occurs within the trapping model that is considered  here \cite{Orenstein1982, Hartenstein1996}; essentially in each case the deep energy states lead to dispersive transport \cite{Borsenberger1992}.
In general the mutiple-trapping WTD is written as \cite{Pollak1977, Jakobs1993}
$\psi(t) = \int_{-\infty}^0 d\epsilon g(\epsilon) k(\epsilon) \exp(-k(\epsilon) t),$
where $g(\epsilon)$ is the density of states and $k(\epsilon) = k_0 e^{\epsilon/k_B T}$.
Hartenstein \emph{et al.} showed \cite{Hartenstein1996} that this multiple-trapping WTD agrees well with the Monte-Carlo-simulated hopping WTD.

\begin{figure}[ptbh]
 \begin{centering}\label{tab:table}
        \includegraphics[scale = 0.365,trim = 130 225 260 175, angle = -0,clip]{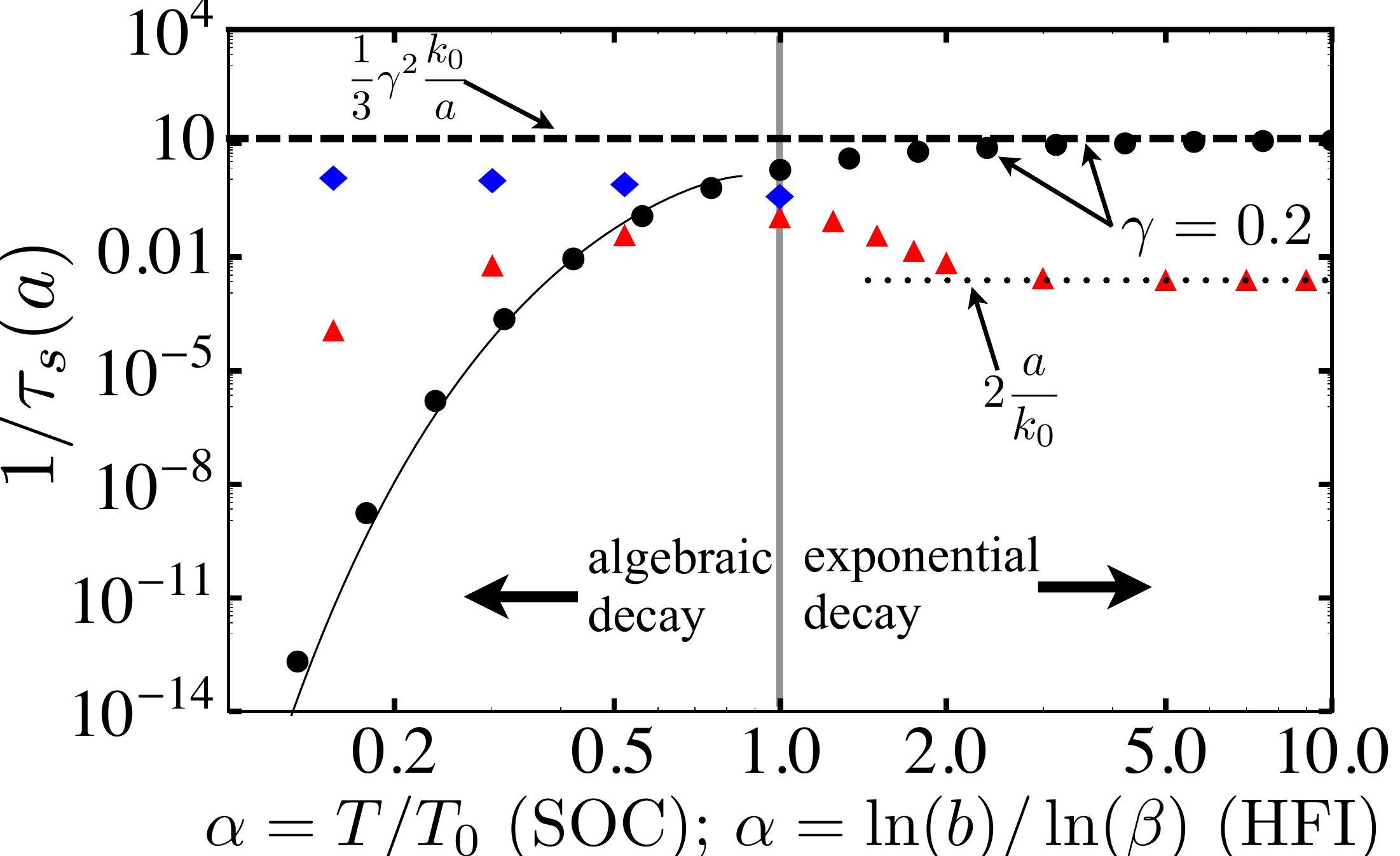}
        \caption[]
{Numerical SOC spin relaxation rate (black solid symbols) and HFI spin relaxation rate (red triangular and blue diamond symbols) as a function of $\alpha$. 
SOC calculations use multiple trapping WTD with exponential DOS while HFI uses the multi-exponential WTD. 
Solid black line is fit using Eq. (\ref{eq:dispEq}) with one multiplicative fit parameter.
Blue diamonds depict the short time spin relaxation. 
Red triangles represent the longer time slower spin relaxation.
Vertical gray line separates regions of algebraic and exponential spin decay. 
$k_0/a = 1000$, $\beta = 0.1$, and $\Gamma = 0$.}\label{fig:dispersiveRelaxation}
        \end{centering}
\end{figure}

At present we consider only an exponential density of states, and note that the end results are also valid for a wide Gaussian density of states ($\alpha$ must be changed to 1/2);  it is advantageous to write the WTD in Laplace space:
\begin{equation}
\tilde\psi(s) = \int_{-\infty}^0 d\epsilon g(\epsilon) \frac{k(\epsilon)}{s+k(\epsilon)} =  \int_{-\infty}^0 dx~ e^{x} \frac{k_0 e^{x/\alpha}}{s+k_0 e^{x/\alpha}},
\end{equation}
where $x = \epsilon/k_B T_0$. The long time behavior of the distribution is algebraic $\psi(t) \sim t^{-1-\alpha}$ where $\alpha = T/T_0$ \cite{Pollak1977} (here $T$ is the temperature and $T_0$ is a parameter characterizing the bandwidth of the states involved in transport).
For SOC, the integral can be computed so the inversion techniques of Ref. \onlinecite{wolframWebsite} can be utilized to solve Eq. (\ref{eq:main}).
However, the form of the spin relaxation for both SOC and HFI can be deduced from the following physical arguments.
From continuous-time random-walk theory \cite{Klafter2011}, the number of hops scales as $n \sim (k_0 t)^{\alpha}$. 
Unsurprisingly for SOC, the polarization falls with increasing number of hops; it can only do so given spin-flips which happen with probability $\sim \gamma^2$. So in general $P(t) \sim 1/(k_0 t \gamma^{2/\alpha})^{\alpha}$ for a heavy-tailed WTD with SOC.
The effect of HFI is more subtle. When hopping is slow ($k_0 < a$), even free hops lead to rapid spin randomization - on the order of the hopping time and independent of the strength of the HFI.
When hopping is fast and traps ineffectual, normal hops ($\alpha> 1 $ in Figure \ref{fig:dispersiveRelaxation}) lead to very little polarization loss since motional narrowing occurs.
When deep traps are present ($\alpha < 1$) then fast hopping entails that deep traps are quickly populated; the resulting polarization behaves similarly to the Kubo-Toyabe polarization function \cite{Kubo1966}. 
This behavior is characterized by two spin relaxation times: at short times, $\tau_s a \sim 1$ (blue diamonds in Figure \ref{fig:dispersiveRelaxation}), and at long times a much longer relaxation time exists (red triangles) that approaches zero as the trap density increases. 
The fast relaxation at short times is strikingly different than free hopping in that the rate is independent of $k_0$.

This reasoning can be proved for SOC - the long time polarization is determined analytically to be algebraically decaying as $P(t) \sim t^{-\alpha}$ \cite{supp}.
The spin relaxation rate (rate to decay to $1/e$) is 
\begin{equation}\label{eq:dispEq}
\frac{1}{\tau_s} \propto \Bigg\{
 \begin{array}{cc}
k_0 \gamma^{2 /\alpha} e^{-1/\alpha} f^{-1/\alpha}& \quad \alpha < 1  \\
k_0 \gamma^2  &\quad \alpha >1  \end{array} 
\end{equation}
and is also displayed in Figure \ref{fig:dispersiveRelaxation} (solid black line) where the role of disorder is apparent.
$f$ has further $\alpha$ dependence but is weak when $\alpha$ not near one.
Since $T_0$ is a measure of the disorder, larger disorder increases $\tau_s$ which agrees with our single trap analysis.
Alternatively, this analytic result can be checked for the exponential density of states and another heavy-tailed WTD (multi-exponential) by finding where the numerical polarization functions decay to $1/e$ (with $\alpha$ defined appropriately). 
The HFI within the multiple-trapping picture poses a much more difficult problem analytically and numerically. 
However it is quite accessible when the multi-exponential WTD is utilized. 
By analogy with SOC, those results are carried over to the multiple-trapping picture with the exponential density of states.

For dispersive transport, a generalized diffusion coefficient is defined as $K_{\alpha} = \overline{r}^2  k_0^{\alpha}/6\Gamma(1+\alpha)$  where $\overline{r}$ is the root-mean-square hopping distance \cite{Klafter2011}. 
This leads to $\ell_s = \sqrt{6 K_{\alpha} \tau_s^{\alpha}}$.
In the non-dispersive regime, the diffusion coefficient is $D = \overline{r}^2 k_0/6$ which entails $\ell_s = \sqrt{6 D \tau_s}$.
As a confirmation of our method, we remark that $\ell_s$ (when using a short relaxation timescale) for fast-hopping-HFI agrees with the analysis of Ref. \onlinecite{Bobbert2009} for the special case of $\alpha=1/2$.
One remarkable feature is that SOC's $\ell_s$ has the same form whether the transport is dispersive or non-dispersive. 
Whereas the SOC spin relaxation rates decrease due to multiple trapping, mobility decreases as well so it is not obvious whether $\ell_s$ increases or decreases in the dispersive compared to the non-dispersive regime.
We find that the $\ell_s$ does not change significantly across the two regimes so that Ref. \onlinecite{Yu2011,Yu2012}'s Efros-Shklovskii hopping prescription for Alq$_3$ \cite{Drew2009} remain valid.
Slow hopping HFI relaxation does not explain the data since the measured $\ell_s$ is always greater than the typical hopping length. 
Alq$_3$  is somewhat unique in that its SOC is large ($\gamma \approx 0.03$  \cite{Yu2012}) compared to other frequently-studied organic semiconductors. 
For instance SOC in MEH-PPV is ~100 times weaker than in Alq$_3$  \cite{Yu2012}.
This explains why  HFI has a large effect on spin preservation in the polymer DOO-PPV \cite{Nguyen2010}.
When deuterated, the spin relaxation rate change is nearly described by the relation found here: $\tau_s^{-1} \propto a$ where $a \approx 0.1 - 0.6$ rad/ns$\cdot$mT \cite{supp}. 
The change in $\ell_s$ also agrees if $\alpha$ is taken to be near unity; changes in $\ell_s$ are more difficult to gauge since it is not a strong function of $a$.

We conclude, by examination of the elements of Table I, that the spin relaxation mechanism in assorted organic semiconductors can be probed by altering the HFI (through deuteration), the hopping rate (through temperature), and SOC (through molecule choice).
This theory should provide a framework for future experimentalists to use for determining spin processes in organic semiconductors. 
In addition, the theory should apply to other disordered systems in which transport is dispersive, such as amorphous inorganic semiconductors and colloidal quantum dot films \cite{GuyotSionnest2007}.

This work was supported by an ARO MURI. 
NJH thanks David Berman for valuable discussions concerning asymptotic expansions.

\end{document}